\documentstyle[eps]{mn}

\newif\ifAMStwofonts

\def\etal{\hbox{\it et al.}}

\ifoldfss
  \ifCUPmtlplainloaded \else
    \NewTextAlphabet{textbfit} {cmbxti10} {}
    \NewTextAlphabet{textbfss} {cmssbx10} {}
    \NewMathAlphabet{mathbfit} {cmbxti10} {} 
    \NewMathAlphabet{mathbfss} {cmssbx10} {} 
  \fi
  \ifAMStwofonts
    \ifCUPmtlplainloaded \else
      \NewSymbolFont{upmath} {eurm10}
      \NewSymbolFont{AMSa} {msam10}
      \NewMathSymbol{\upi}     {0}{upmath}{19}
      \NewMathSymbol{\umu}     {0}{upmath}{16}
      \NewMathSymbol{\upartial}{0}{upmath}{40}
      \NewMathSymbol{\leqslant}{3}{AMSa}{36}
      \NewMathSymbol{\geqslant}{3}{AMSa}{3E}

      \let\leq=\leqslant 
      \let\geq=\geqslant 
    \fi
  \fi
\fi 

\ifnfssone
  \newmathalphabet{\mathit}
  \addtoversion{normal}{\mathit}{cmr}{m}{it}
  \addtoversion{bold}{\mathit}{cmr}{bx}{it}
  \newmathalphabet{\mathbfit} 
  \addtoversion{normal}{\mathbfit}{cmr}{bx}{it}
  \addtoversion{bold}{\mathbfit}{cmr}{bx}{it}
  \newmathalphabet{\mathbfss} 
  \addtoversion{normal}{\mathbfss}{cmss}{bx}{n}
  \addtoversion{bold}{\mathbfss}{cmss}{bx}{n}
  \ifAMStwofonts
    \ifCUPmtlplainloaded \else
      \UseAMStwoboldmath
      \makeatletter
      \new@mathgroup\upmath@group
      \define@mathgroup\mv@normal\upmath@group{eur}{m}{n}
      \define@mathgroup\mv@bold\upmath@group{eur}{b}{n}
      \edef\UPM{\hexnumber\upmath@group}
      \new@mathgroup\amsa@group
      \define@mathgroup\mv@normal\amsa@group{msa}{m}{n}
      \define@mathgroup\mv@bold\amsa@group{msa}{m}{n}
      \edef\AMSa{\hexnumber\amsa@group}
      \makeatother
      \mathchardef\upi="0\UPM19
      \mathchardef\umu="0\UPM16
      \mathchardef\upartial="0\UPM40
      \mathchardef\leqslant="3\AMSa36
      \mathchardef\geqslant="3\AMSa3E

      \let\leq=\leqslant 
      \let\geq=\geqslant 
    \fi
  \fi
\fi 

\ifnfsstwo
  \DeclareMathAlphabet{\mathbfit}{OT1}{cmr}{bx}{it}
  \SetMathAlphabet\mathbfit{bold}{OT1}{cmr}{bx}{it}
  \DeclareMathAlphabet{\mathbfss}{OT1}{cmss}{bx}{n}
  \SetMathAlphabet\mathbfss{bold}{OT1}{cmss}{bx}{n}
  \ifAMStwofonts
    \ifCUPmtlplainloaded \else
      \DeclareSymbolFont{UPM}{U}{eur}{m}{n}
      \SetSymbolFont{UPM}{bold}{U}{eur}{b}{n}
      \DeclareSymbolFont{AMSa}{U}{msa}{m}{n}
      \DeclareMathSymbol{\upi}{0}{UPM}{"19}
      \DeclareMathSymbol{\umu}{0}{UPM}{"16}
      \DeclareMathSymbol{\upartial}{0}{UPM}{"40}
      \DeclareMathSymbol{\leqslant}{3}{AMSa}{"36}
      \DeclareMathSymbol{\geqslant}{3}{AMSa}{"3E}

      \let\leq=\leqslant 
      \let\geq=\geqslant 
    \fi
  \fi
\fi 

\ifCUPmtlplainloaded \else
  \ifAMStwofonts \else 
    \def\upi{\pi}
    \def\umu{\mu}
    \def\upartial{\partial}
  \fi
\fi

\title{Weighing a galaxy bar in the lens Q2237+0305}
\author[Schmidt, Webster \& Lewis]  
{Robert Schmidt,$^1$\thanks {Present address:
Astro\-phy\-si\-ka\-li\-sches Insti\-tut Potsdam, An der
Stern\-war\-te 16, 14482 Pots\-dam, Ger\-many. E-mail:
rschmidt@\protect\linebreak[0]aip.de}
Rachel L. Webster$^1$ and 
Geraint F. Lewis$^2$\thanks {Visiting Astronomer at the University of
Melbourne}\\
$^1$ School of Physics, University of Melbourne, Parkville, Victoria 3052, 
Australia.\\
$^2$ SUNY at Stony Brook, Stony Brook, NY11794-2100, USA.\\}

\date{}

\pagerange{\pageref{FirstPage}--\pageref{LastPage}}
\pubyear{1996}

\begin{document}

\maketitle

\label{FirstPage}

\begin{abstract}
In the gravitational lens system Q2237+0305 the cruciform quasar image
geometry is twisted by ten degrees by the lens effect of a bar in the
lensing galaxy. This effect can be used to measure the mass of the
bar. We construct a new lensing model for this system with a power-law
elliptical bulge and a Ferrers bar. The observed ellipticity of the
optical isophotes of the galaxy leads to a nearly isothermal
elliptical profile for the bulge with a total quasar magnification of
$16^{+5}_{-4}$. We measure a bar mass of $7.5\pm 1.5\times10^{8}\,{\rm
h}_{75}^{-1}{\cal M}_{\sun}$ in the region inside the quasar images.
\end{abstract}


\section{Introduction}

Gravitational lensing provides a unique way to weigh objects at
cosmological distances without any assumption about the connection
between light and dark matter. Since the discovery of the first
gravitational lens (Walsh, Carswell \& Weymann 1979) several
gravitational lenses have been found and this method has been used
many times to explore the mass distribution of galaxies. In this
paper, we model the lens system Q2237+0305 in order to weigh the bar
in the lensing galaxy.

The quasar Q2237+0305 ($z_{q} = 1.695$) was found by Huchra {\etal}
\shortcite{Huchra1985} at the centre of an SBb spiral galaxy
($z=0.0394$) that is situated in the outskirts of the Pegasus~II
cluster. The quasar was later resolved into four images that are
situated around the core of the galaxy within a radius of one
arcsecond \cite {Yee1988,Schneider1988}.

Two fundamentally different approaches have been used to model the
lensing galaxy. One was to fit a parametric mass profile with several
free parameters to the observed quasar image
configuration~\cite{Kent1988}; the other to use a model of the light
distribution of the galaxy and to fit for the mass-to-light ratio as
the single free parameter \cite {Schneider1988,Rix1992}. The former
approach was naturally much more precise in the reproduction of the
observed image geometry due to the greater number of free parameters.

The length scale over which the lensing galaxy influences a light
bundle from
the quasar is small compared to the cosmological distances between
observer, lens and source. The lens can therefore be treated as a mass
sheet at the position of the galaxy. Since the galaxy disk of
2237+0305 is inclined with respect to the sky, elliptical surface mass
distributions must be used in the models of this system.

\begin{figure}
\plotone{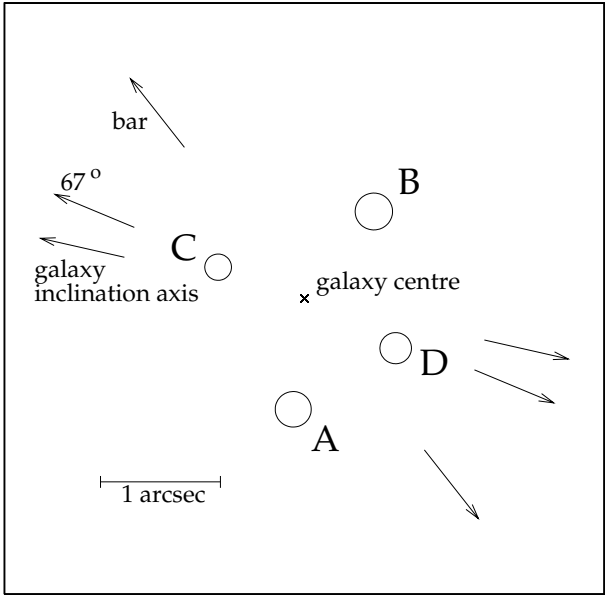}
\caption{Illustration of the image geometry of Q2237+0305, motivated
by figure~1 in Kent \& Falco \protect\shortcite {Kent1988}. The images
are labelled using the convention by Yee \protect\shortcite {Yee1988}
with positions from Crane {\etal} \protect\shortcite {Crane1991}. The
relative areas of the circles correspond to the radio flux ratios from
Falco {\etal} \protect\shortcite {Falco1996}. The position of the
galaxy centre is indicated with a cross. The long arrows indicate the
directions of the galaxy inclination axis and the bar, as well as the
position angle of $67\degr$. North is up, East to the left.}
\label{ImageGeometry}
\end{figure}

Interestingly, the position angle, counted counterclockwise from north,
of the major axis of the elliptical lens models found by Kent \& Falco
\shortcite {Kent1988} was about $67\degr$. This is almost parallel to
the axis through images C and D and just between the angle of the
inclination axis of the galaxy ($77\degr$, Yee 1988) and the angle of
the bar ($39\degr$, also Yee 1988). This situation is shown in
figure~\ref {ImageGeometry}. This was also found by Kochanek \shortcite {Kochanek1991}
and Wambsganss \& Paczy\'{n}ski \shortcite {Wambsganss1994} who used
simple circular mass distributions with an additional quadrupole
perturbation. When fitted to the observed image geometry, the
direction of the perturbation turned out to be close to the one Kent
\& Falco \shortcite {Kent1988} found for their model major axis. More
recent investigations by Witt, Mao \& Schechter \shortcite {Witt1995},
Kassiola \& Kovner \shortcite {Kassiola1995} and Witt \shortcite
{Witt1996} obtained the same position angle for the perturbation or
major model axis.

On the other hand, the bar shows up prominently in CCD images of the
galaxy. It has been noted~\cite {Tyson1985,Yee1988,Foltz1992} that it
might contribute significantly to the lensing in the system --
initially this was actually an aid to explain the lensing effect when
only the images A, B and C were known~\cite {Tyson1985}.

The motivation for this paper is the idea that the apparent
misalignment of the predicted model major axis and observed galaxy
inclination axis as shown in figure~\ref {ImageGeometry} is due to the
lensing influence of the bar. In section~\ref {TheoreticalModel} we
construct and analyse a lensing model that includes the bar component
and takes the observed position angle for the inclination axis of the
galaxy into account. Section~\ref {PropertiesOfTheBar} deals with the
implications of this model for the bar. In section~\ref {Discussion}
we finally discuss our results. We use a cosmological model with $H_0
= 75\,{\rm h_{75}}\, {\rm km\, s^{-1}\, Mpc^{-1}}$, $\Omega=1$ and
$\Lambda=0$.

\section{Theoretical Model}

\label{TheoreticalModel}

The lensing model we construct has two components, each with several
free parameters. In this section, we introduce the components and
determine the values for these that provide the best description of
the observations.

\subsection{The bulge}
\label{TheBulge}

Yee \shortcite {Yee1988} identified three components in the inner part
of the galaxy: bulge, disk and bar. In the lensing models for this
system, bulge and disk have been represented by just one effective
component since there is only limited observational data from the
quasar image and galaxy positions to constrain free parameters of the
model. For simplicity, we call this composite component 'bulge'.

Moreover, Kochanek \shortcite {Kochanek1991} and Wambsganss \&
Pa\-czy\'{n}ski \shortcite {Wambsganss1994} found that the quasar
image positions and the galaxy position in this system do not
constrain the parameters even of simple lens models. In particular,
Wambsganss \& Paczy\'{n}ski \shortcite {Wambsganss1994} showed that
for a circular power--law mass distribution with an external shear
there is a whole family of models that fit the observations; they
discovered that for this family there is a linear relation between the
magnitude of the external shear and the exponent of the mass profile
for a vast range of exponents. The shear and the mass exponent are
degenerate and one needs more information than only the positions of
the quasar images and the galaxy to break this degeneracy. If one uses
a two-component galaxy model the shear contributions from the two
components will also be degenerate since the resulting shear is
degenerate.

One way to get around the shear degeneracy is to use an elliptical
mass distribution instead of a circular profile. In this case, the
ellipticity--parameter of the mass distribution replaces the
shear--parameter as a free parameter of the model. An analogous
degeneracy in the ellipticity can then be broken by using the observed
ellipticity of the isophotes of the galaxy.

Generalising the approach by Wambsganss \& Paczy\'{n}ski \shortcite
{Wambsganss1994}, we accordingly modelled the bulge with an elliptical
power--law mass distribution with a major axis position angle of $77
\degr$. There is some disagreement in the literature on the value of
this position angle (for example Rix, Schneider \& Bahcall
1992). Fitte \& Adam \shortcite {Fitte1994} showed that this is
because the position angle of elliptical isophotes is increasingly
twisted towards the bar with increasing distance from the galaxy
centre. The value we adopted from Yee \shortcite {Yee1988} is
identical with the position angle determined from the galaxy continuum
map within a radius of one arcsecond from the galaxy centre \cite
{Fitte1994} where most of the lensing mass is situated. Let $\epsilon$
be the elliptical parameter, so that the ratio $b/a$ of minor and
major axis of concentric elliptical shells is

\begin{equation}
{b\over a} = \frac{1-\epsilon}{1+\epsilon}.
\end{equation}

\noindent It is useful to express surface mass densities in units of
the critical lensing density \cite {Schneider1992} $\Sigma_{\rm
crit}=c^2D_{\rm s}/4\pi {\rm G}D_{\rm d}D_{\rm ds}$, where $D_{\rm s}$,
$D_{\rm d}$ and $D_{\rm ds}$ are the angular size distances between
observer and source, observer and deflector (lens), as well as
deflector and source. The surface mass density $\kappa$ of a
power--law elliptical mass distribution in units of \mbox{$\Sigma_{\rm
crit}$} is given by

\begin{equation}
\label{SingularEllipticalProfiles}
\lefteqn{\kappa (\theta_1,\theta_2) = \frac{E_{0}}{2\,\theta_e^{\nu}}.}
\\
\end{equation}

\noindent $\theta_1$ and $\theta_2$ are the coordinates on the sky as
measured in a coordinate system oriented with the observed major and
minor axis of the bulge. $\theta_e$ is the elliptical radius

\begin{equation}
\theta_e=\sqrt{\frac{\theta_1^{2}}{(1+\epsilon )^{2}} +
\frac{\theta_2^{2}}{(1-\epsilon)^{2}}},
\end{equation}

\noindent $\nu$ is the power--law exponent of the elliptical mass
distribution and $E_{0}$ is a constant. In the analysis of our results
we use the ellipticity $e=1-\frac{b} {a}=\frac {2\epsilon}
{1+\epsilon}$ since this is the value that is usually used in the
observations. In Appendix~\ref {PowerLawDeflection} deflection
potential and angles for this mass distribution are described.

\subsection{The bar}
\label{TheBar}

The light distribution of bars has a well-defined elongated shape, and
is non-singular and centrally condensed \cite{Sellwood1993}. It is not
straightforward to determine the true form of bar mass distributions
from this, so that we have to assume a model. Very simple models with
these properties of the bar light are the Ferrers profiles \cite
{Ferrers1877}. They were used in dynamical studies of bars \cite
{Freeman1966,Martinet1988,Sellwood1993} since they can be treated
analytically. To model the surface mass distribution of the bar of
2237+0305 we used two-dimensional Ferrers profiles of the form

\begin{equation}
\label{FerrersProfiles}
\kappa (\theta_1,\theta_2) = \left\{ \begin{array}{ll} \kappa_{\rm c}
         {(1-\frac{\theta_1^{2}}{a^2}-\frac{\theta_2^{2}}{
         b^2})}^{\lambda} & \mbox{if $\frac{\theta_1^{2}}{
         a^2}+\frac{\theta_2^{2}}{b^2} \leq 1$} \\ 0 &
         \mbox{otherwise} \end{array} \right. \!\!\!\!.
\end{equation}

\noindent In this equation, $\theta_1$ and $\theta_2$ are the
coordinates on the sky as measured in a coordinate system oriented
with the observed major and minor axis of the bar.  $\lambda$ is a
real number, $\kappa_{\rm c}$ is the central surface density in units
of the critical lensing density $\Sigma_{\rm crit}$ defined in
section~\ref {TheBulge} and $a$, $b$ are the semi-major respectively
semi-minor axis of the bar.

In the analysis we restricted ourselves to moderate exponents
$\lambda=0.5$, $1$ and $2$. The deflection potential and angles for
integer values of $\lambda$ can be calculated analytically as
described in Appendix~\ref {FerrersDeflection}. For $\lambda=0.5$ the
profiles were constructed through the numerical superposition of many
elliptical slices of constant density ($\lambda=0$) and different
size \cite {Schramm1994}.

\subsection{The effect of shear}
\label{Theeffectofshear}

The lensing influence of the bar can be understood by considering a
system with two shear tensors with shear directions as shown in
figure~\ref {ImageGeometry} for galaxy inclination axis and bar (see
Schneider, Ehlers \& Falco \shortcite {Schneider1992} for the
definition of the shear tensor). The resulting shear tensor can be
found by adding up the single tensors and the resulting shear
direction is determined by the shear ratio of the two components.

A source almost directly behind the core of the lensing galaxy appears
lensed with four of the five images in a cross formation aligned with
the axes parallel and perpendicular to the resulting shear direction,
while the fifth is seen in the centre \cite [p252]{Schneider1992}. In
the case of 2237+0305, the major axis position angle as found by the
one-component lens models ($67\degr$) can be interpreted to be this
resulting shear direction, which almost coincides with the axis
through images C and D. The axis through images C and D has
effectively been twisted away from the galaxy inclination axis by the
bar.

\begin{figure}
\plotone{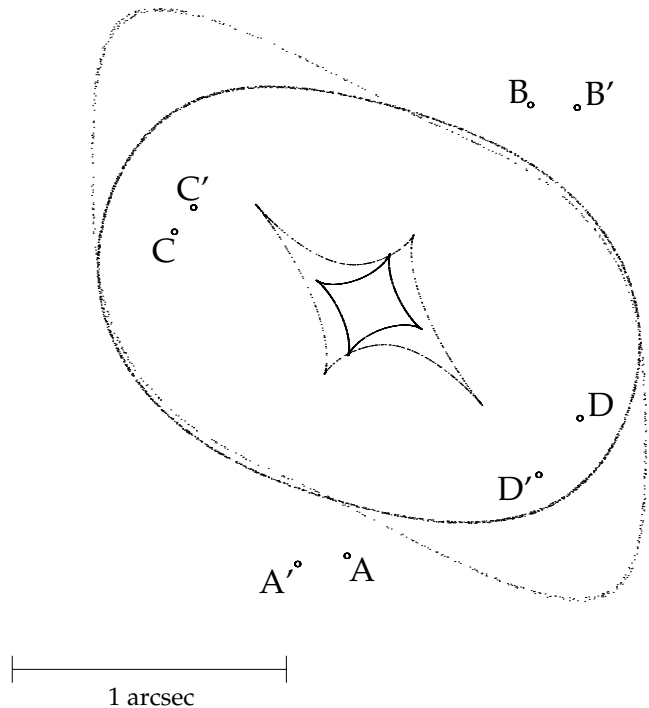}
\caption{Illustration of the effect of a strong bar. Assuming
identical source positions, the diamond-shaped caustics, the
corresponding critical curves and the image positions are plotted for
two different lenses. The smaller caustic and the almost elliptical
critical curve belong to a model with a $\nu=1$ power--law bulge and a
$\lambda =2$ Ferrers bar that was fitted to the observed parameters of
Q2237+0305.  The positions of the images created by this model are
labelled with unprimed letters as in figure~\protect\ref
{ImageGeometry}. In this model, the bulge mass inside the ring of
images is about 20 times larger than the bar mass (see
table~\protect\ref {ParametersofModels}). The caustic and the critical
curve transform into the two elongated curves if the mass of the bar
inside the ring of images is increased to half the mass of the bulge,
while the bulge mass is kept fixed. The images are shifted by the more
massive bar to the positions labelled with the primed letters.}
\label{StrongBar}
\end{figure}

In figure~\ref {StrongBar} this twisting is illustrated by plotting
the critical lines, caustics and image positions of two barred lenses
with identical source positions.  The caustics are the lines in the source plane that separate
regions of different image multiplicity. The critical lines are the
corresponding lines in the lens plane where pairs of images are
created or destroyed \cite {Schneider1992}.

In this figure, the first lens has a weak bar that barely changes the
elliptical shape of the bulge's critical line or the corresponding
diamond shape of the caustic. The other bar is significantly more
massive; it warps the shape of these structures and shifts the image
positions.

\subsection{Detailed modelling}
\label{DetailedModelling}

Our barred galaxy model has ten adjustable parameters. These are the
positions of the galaxy and the source plus the constant $E_0$, the
ellipticity $e$ and the exponent $\nu$ for the bulge as well as the
bar mass normalisation $\kappa_{\rm c}$, the semi-minor axis $b$ and
the exponent $\lambda$ for the bar. This number can be reduced by two
if the observed ellipticity of the bulge and only fixed values for
$\lambda$ are used. For the length of the semi-major axis of the bar
we used the observed value of \mbox{$a \approx 9$ arcsec} (taken from
the figures in Yee 1988 or Irwin {\etal} 1989). The lens effect is
insensitive to the precise value of $a$ because $a$ is much larger
than the radius of the ring of images ($\approx$~1~arcsec). The length
of the semi-minor axis must, however, remain a free parameter of the
model since it is comparable to this radius, but not known well
enough (\mbox{${\rm b}\approx 1-2$ arcsec}, see section~\ref
{PropertiesOfTheBar}).

There are ten observational constraints the system imposes upon
theoretical models. These are the coordinates of the four observed
quasar images and the galaxy centre. The positions for the images and
galaxy centre were taken from Crane {\etal} \shortcite
{Crane1991}. These positions have been determined from {\it Hubble
Space Telescope (HST)} observations and have quoted measurement errors
of $0\farcs 005$. In general, the ratios of the fluxes of the
different images of a gravitational lens also provide good constraints
for a model. Unfortunately, in the case of Q2237+0305 the lightcurves
from Corrigan {\etal} \shortcite {Corrigan1991} or {\O}stensen {\etal}
\shortcite {Ostensen1996} clearly show that all optical image fluxes
are subject to flux variations due to microlensing of the quasar light
from the stars in the lensing galaxy. In addition, the light from the
quasar is non-uniformly dust reddened during the passage through the
galaxy. The fluxes were, therefore, neglected in the modelling
procedure. We will, however, compare the model predictions with the
recently measured radio flux densities by Falco {\etal} \shortcite
{Falco1996}.

Let $\vec{\btheta}_{k}$, $\vec{\btheta}_{g}$ be the positions on the
sky the model predicts for quasar images and the galaxy centre and
$\vec{\btheta}_{ko}$, $\vec{\btheta}_{go}$ the observed positions with
their positional uncertainties $\sigma_{k}$, $\sigma_{g}$. To find the
best fit model, the expression

\begin{equation}
\label{ChisquareWP}
\chi^{2} = \sum_{k=1}^{4} \frac {{\left( \vec{\btheta}_{k}- \vec
{\btheta}_{ko} \right)}^ {2}}{\sigma^{2}_{k}} + \frac {\left(
\vec{\btheta}_{g}- \vec{\btheta}_{go} \right)^2} {\sigma^{2}_{g}}
\end{equation}

\noindent \cite {Wambsganss1994} was minimised through variation of
the model parameters using a multidimensional minimisation routine
(direction set or downhill simplex methods according to Press {\etal}
1992).

In order to find an estimator for the separations $\vec{\btheta}_{k}-
\vec {\btheta}_{ko}$ between modelled and observed images for the
first term on the right hand side of Equation~(\ref {ChisquareWP}), we
used the method by Kochanek \shortcite {Kochanek1991}; the separations
between an optimally weighted source position and the positions in the
source plane where the observed image positions are mapped to by a
given lens model are propagated back into the lens plane.

\subsection{Analysis}
\label{Analysis}

For a nearly circularly symmetric lens with a source almost in the
origin, it follows from Newton's theorem in two dimensions \cite
{Foltz1992,Schramm1994} that the mass inside the circle of images is
approximately given by the separation $\Delta \theta$ of the images at
opposite ends of the cross via

\begin{equation}
\label{EinsteinMass}
M=\frac{c^2}{16\,{\rm G}} \frac{D_{\rm d} D_{\rm s}}{D_{\rm ds}}
\left(\Delta \theta \right)^2
\end{equation}

\noindent (see for example Narayan \& Bartelmann 1996). The galaxy
2237+0305 is situated relatively close to us at an angular size
distance
\mbox{$D_{\rm d}=0.15\, {\rm h}_{75}^{-1}$ Gpc}, so that $D_{\rm
ds}/D_{\rm s}\approx 1$. The separations of the quasar images are
\mbox{$\Delta \theta\approx 1.8$} arcsec, so that we get
$M\approx1.5\times10^{10}\,{\rm h}_{75}^{-1}{\cal M}_{\sun}$. This
value was also found in previous models for this system (Rix
{\etal}~1992: $1.44\pm 0.03\times 10^{10} {\rm h}_{75}^{-1} {\cal
M}_{\sun}$, Wambsganss \& Paczy\'{n}ski~1994: $1.48\pm 0.01\times
10^{10} {\rm h}_{75}^{-1} {\cal M}_{\sun}$). The other value
theoretical models for 2237+0305 agreed on was the resulting shear
direction of~$\approx 67\degr$.

In our barred lens model, the bulge acts as the main lensing mass and
the bar as a perturbation; for a given ellipticity $e$ or exponent
$\nu$, the bulge parameter $E_0$ and hence
the bulge mass do not change very much for different $\lambda$-bar
models. In fact, experiments with different bar masses as in
figure~\ref {StrongBar} showed that for similar masses of bulge and
bar inside the quasar images the cruciform image symmetry gets
skewed. Models with a strong bar thus cannot reproduce a symmetric
image geometry as in Q2237+0305. In order to explore the effect of
$\lambda$ on the models, we used fixed values $\lambda=0.5$, $1$ and
$2$.

\begin{figure} \plotone{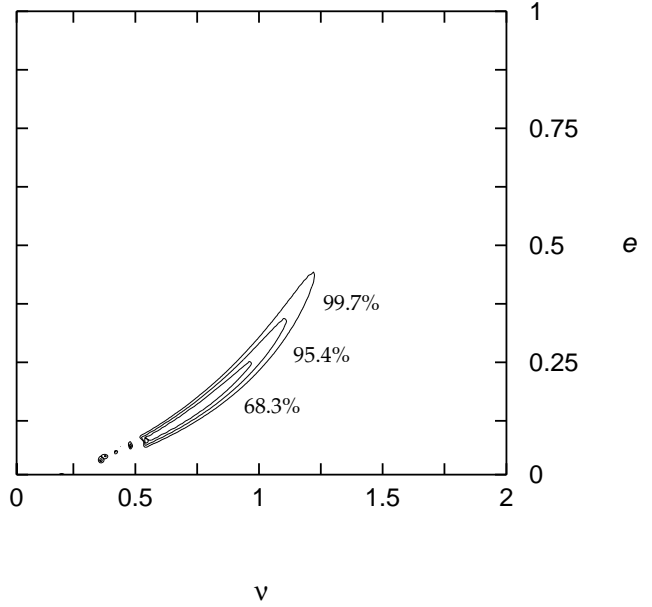}
\caption{Contour plot of confidence regions in the parameter space of
bulge exponent $\nu$ and ellipticity $e$. The indicated contours
contain $68.3\%$, $95.4\%$ and $99.7\%$ of normally distributed models
around the minimum of~$\chi^2$.}
\label{enuDegeneracy}
\end{figure}

Besides $E_0$ and $\lambda$, the parameter space of $e$, $\nu$,
$\kappa_{\rm c}$ and $b$ had to be examined. We first scanned the
parameter space of $e$ and $\nu$ while leaving $\kappa_{\rm c}$ and
$b$ as free parameters. In figure~\ref {enuDegeneracy} a contour plot
of the confidence regions that contain 68.3\%, 95.4\% and 99.7\% of
normally distributed models around the minimum of $\chi^2$ \cite
{Press1992} in the parameter space of $\nu$ and $e$ is
shown.  Only the
plot for $\lambda=2$ is presented; the cases with $\lambda=0.5$ and
$\lambda=1$ are very similar. The parameter space was scanned with a
stepsize of $0.02$ for $\nu$ and $0.01$ for $e$. For every point the
best parameters were determined through minimisation and a
$\chi^2$-value was computed. The best fit models lie in a long valley
that extends up to $\nu\approx 1.25$. The unclear structure at the
lower end of the valley for $\nu\leq 0.5$ and $e\leq 0.1$ is due to
numerical effects. Detailed investigation shows that the valley
continues towards smaller $\nu$, becoming shallower. The models in
this region, with low $e$ and $\nu$, are similar to circular disks
with constant surface mass density and are not examined here because
they do not represent realistic galaxy models.

The external-shear models by Wambsganss \& Pa\-czy\'{n}ski \shortcite
{Wambsganss1994} yielded no constraints on their circular mass
distributions for the whole range of parameters from \mbox{$\nu=0.07$}
to \mbox{$\nu=2$}. The smaller allowed range for $\nu$ in figure~\ref
{enuDegeneracy} shows that the ellipticity does not allow the same
kind of freedom as the shear. For high ellipticities of the power--law
bulges, the quasar image geometry cannot be reproduced anymore.

\begin{figure}
\plotone{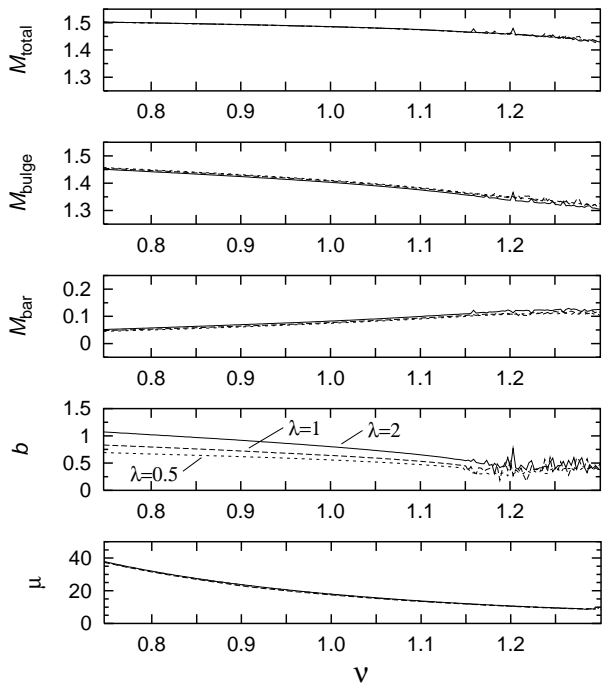}
\caption{Model parameters along the valley of best fits. Plotted
against $\nu$ are the total mass and the masses of bulge and bar
inside a circle of $0.9$ arcsec in~$10^{10}\,{\rm h}^{-1}_{75}\,{\cal
M}_{\odot}$ as well as the semi-minor bar axis $b$ in arcsec and the
total magnification $\mu$. Different line styles have been used for
different values of $\lambda$ as indicated in the $b$-$\nu$ panel.}
\label{ModelParameters}
\end{figure}

In figure~\ref {ModelParameters} the total mass, the bulge mass, and
the bar mass inside a circle of $0.9$ arcsec as well as the semi-minor
axis $b$ and the total magnification $\mu$ along the valley of best
fits around the $\nu=1$ profile are shown for $0.75 \leq \nu \leq 1.3$
and the three values of $\lambda$. 
The plots for different $\lambda$ only differ noticeably in their
semi-minor axis predictions. Beginning with $\nu\approx 1.2$ the minimisation
produces numerical noise at the upper end of the valley from
figure~\ref {enuDegeneracy} since the fits get worse.

It can be seen that the total mass of the model inside a circle of
$0.9$ arcsec is almost constant. The constituent masses of the bar and
the bulge exhibit a dependence on $\nu$. It can be interpreted that
the bar mass increases with $\nu$ in order to counter the increased
contribution of the bulge to the resulting shear due to the increase
of $e$, and hence the bulge mass decreases in order to conserve the
mass inside $0.9$ arcsec given by equation~(\ref {EinsteinMass}). The
magnification drops strongly with increasing $\nu$, which was also
observed by Wambsganss \& Paczy\'{n}ski \shortcite {Wambsganss1994}.

As motivated in the introduction, we chose the model from the family
of best models in figure~\ref {enuDegeneracy} that exhibits the
observed ellipticity of the bulge. The ellipticity has been measured
by Racine~\shortcite {Racine1991} to be $e=0.31\pm 0.02$, in agreement
with the results from the continuum map by Fitte \& Adam \shortcite
{Fitte1994}. This range of ellipticities is encompassed by the range
$\nu = 1.05\pm 0.1$ of bulge exponents. We use this region of allowed
bulge models to determine the uncertainties of the predictions of
models with different bars.

From all possible values for $\nu$ and the three bar models we obtain
an estimate of the total mass inside a circle of 0.9 arcsec of
$1.49\pm 0.01 \times 10^{10} {\rm h}_{75}^{-1} {\cal M}_{\sun}$. This
is consistent with the estimate and the results from the literature
given at the beginning of this section. In a review, Narayan \&
Bartelmann \shortcite {Narayan1996} mention only two cases of
gravitational lenses in which it has been possible to constrain the
radial distribution of the lens galaxy. Both models are based on
singular elliptical or nearly elliptical profiles. For MG 1654+134,
Kochanek \shortcite {Kochanek1995} obtained $\nu=1.0\pm0.1$, and for
QSO 0957+561, Grogin \& Narayan \shortcite {Grogin1996} obtained
$\nu=1.1\pm 0.1$. The predictions of our model for the bulge are in
good agreement with these values.

In table~\ref {ParametersofModels}, the model parameters for the
different values of $\lambda$ and the uncertainties due to the
uncertainty in $\nu$ are shown. \begin{table*}
\begin{center}
\begin{minipage}{140mm}
\begin{center}
\caption{Model Parameters for Q2237+0305 for three values of the bar
exponent $\lambda$ and the bulge exponent $\nu$. The columns for
$\lambda=0.5$, 1.0 and 2.0 are indicated. Each table entry contains
the value for $\nu=1.05$ and the differences to the corresponding
values for $\nu=1.15$ (upper index) and $0.95$ (lower index). Only one
value is given for a line if the differences between the bar models
are below the rounding precision.  $\chi^2$ is divided by three, the
number of degrees of freedom. $E_0$, $b$ and the source position
$\left( \beta_1, \beta_2 \right)$ are given in arcsec. $\mu_{\rm
total}$ is the total magnification, $\mu_{\rm ij}$ are the relative
magnification ratios between the images. $\Delta t_{\rm ij}$ are the
relative time--delays between the images in ${\rm h}_{75}^{-1}$
hours. $M_{\rm bulge}(<0.9\arcsec)$ and $M_{\rm bar}(<0.9\arcsec$) are
the masses of bulge and bar inside a circle of $0.9$ arcsec and
$M_{\rm bar, total}$ is the total mass of the bar. Masses are given in
$10^{10} {\rm h}_{75}^{-1} {\cal M}_{\odot}$. The semi-major bar axis
was assumed as $a=9$ arcsec. The {\it VLA} flux ratios are the~3.6cm
radio flux ratios measured by Falco {\etal} \protect\shortcite
{Falco1996}.}
\label{ParametersofModels}
\begin{tabular}{@{}lllll@{}} 
& & & & {\it VLA} flux\\ $\lambda$ & 0.5 & 1.0 & 2.0 & ratios \\ \hline
$\chi^ {2}$/3 & & $2.1^{+1.3}_{-0.8}$\\ \\
$E_{0}$ & & $0.81^{-0.09}_{+0.09}$\\
$e$ & & $0.31^{+0.07}_{-0.06}$\\ \\
$\kappa_{\rm c}$ & $0.074^{+0.022}_{-0.014}$ &
$0.079^{+0.021}_{-0.016}$ & $0.085^{+0.023}_{-0.017}$ \\
$b$ & $0.58^{-0.06}_{+0.04}$ & $0.67^{-0.07}_{+0.06}$ &
$0.85^{-0.10}_{+0.08}$ \\ \\
$\beta_1$ & & $-0.063^{-0.010}_{+0.009}$\\
$\beta_2$ & & $-0.014^{-0.003}_{+0.001}$\\ \\
$\mu_{\rm total}$ & $15.6^{-3.5}_{+4.8}$ & $16.0^{-3.6}_{+5.0}$ & 
$16.2^{-3.6}_{+5.0}$ \\
$\mu_{\rm BA}$ & $1.12^{-0.02}_{+0.01}$ & $1.08^{+0.03}_{-0.04}$ &
$1.04^{+0.03}_{-0.03}$ & $1.08\pm0.27$\\
$-\mu_{\rm CA}$ & $0.62^{-0.04}_{+0.04}$ & $0.61^{-0.02}_{+0.01}$ &
$0.60^{-0.02}_{+0.02}$ & $0.55\pm0.21$ \\
$-\mu_{\rm DA}$ & $1.26^{-0.07}_{+0.06}$ & $1.26^{-0.01}_{-0.01}$ &
$1.24^{+0.02}_{\pm0.0}$ & $0.77\pm0.23$\\ \\
$-\Delta {\rm t}_{\rm BA}$ & & $2.0^{+0.4}_{-0.3}$ \\
$\Delta {\rm t}_{\rm CA}$ & & $16.2^{+2.7}_{-4.4}$ \\
$\Delta {\rm t}_{\rm DA}$ & & $4.9^{+1.0}_{-0.8}$ \\ \\
$M_{\rm bulge}(<0.9\arcsec)$ & $1.42^{-0.02}_{+0.01}$ &
$1.42^{-0.02}_{+0.01}$ & $1.41^{-0.01}_{+0.02}$ \\
$M_{\rm bar}(<0.9\arcsec)$ & $0.07^{+0.01}_{-0.01}$ &
$0.07^{+0.01}_{-0.01}$ & $0.08^{+0.01}_{-0.01}$ \\
$M_{\rm bar, total}$ & $0.47^{+0.07}_{-0.06}$ & $0.43^{+0.06}_{-0.06}$
& $0.39^{+0.05}_{-0.05}$ \\ \hline
\end{tabular}
\end{center}
\end{minipage}
\end{center}
\end{table*} The $\chi^2$-values from equation~(\ref {ChisquareWP}) have
been divided by the number of degrees of freedom, giving a 
measure of the quality of the fit. The number of degrees of freedom is
the number of constraints minus the number of free parameters; here we
have three degrees of freedom since $e$, $\nu$ and $\lambda$ are
fixed.

The value for the total magnification predicted from the $\nu=1.05$
model is about half the value found by Wambsganss \& Paczy\'{n}ski
\shortcite {Wambsganss1994} for their $\nu=1$ circular power law
profile with an external shear. This illustrates that the use of an
elliptical mass distribution drastically changes the predictions of
the model; a similar discrepancy is apparent for the time--delays. Note
that $\mu_{\rm total}\approx \frac{3}{\epsilon}=16.7$ with
$\epsilon=0.18$ (corresponds to $e=0.31$, see section~\ref {TheBulge})
as derived for a $\nu=1$ power--law lens with small $\epsilon$ and a
source in the origin by Kassiola \& Kovner \shortcite
{Kassiola1993}. In contrast to the non-singular mass models by Kent \&
Falco \shortcite {Kent1988} our lens models do not produce a fifth
image in the centre due to the central singularity of the mass
distribution. The predicted source positions are very similar to the
ones by Kent \& Falco.

\begin{figure} \plotone{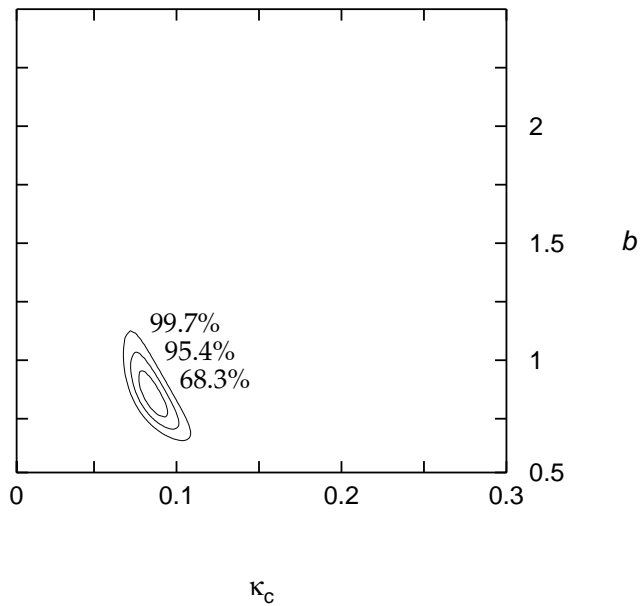}
\caption{Contour plot of confidence regions in the parameter space of
bar normalisation $\kappa_{\rm c}$ and semi-minor axis b (in arcsec)
for a bulge exponent $\nu=1.05$, ellipticity $e=0.31$ and bar exponent
$\lambda=2$. The indicated contours contain $68.3\%$, $95.4\%$ and
$99.7\%$ of normally distributed models around the minimum
of~$\chi^2$.}
\label{kappa0b}
\end{figure}

The remaining parameter space of the bar models is illustrated in
figure~\ref{kappa0b}. Similar to figure~\ref {enuDegeneracy}, the
contours of the confidence regions of normally distributed models
around the minimum of $\chi^2$ in the parameter space of $\kappa_{\rm
c}$ and $b$ are shown for the model with $\nu=1.05$, $e=0.31$
and $\lambda=2$. The parameter space was scanned with a stepsize of
$0.003$ in $\kappa_{\rm c}$ and $0.025$ arcsec in $b$. The bar
parameters are well constrained at the bottom of a steep
boomerang-shaped valley.

The surface mass density $\kappa$ in units of the critical density and
the local shear $\gamma$ at the positions of the images are given in
table~\ref {LocalLensingParameters}.\begin{table}
\begin{center}
\caption{Local lensing parameters at the positions of the quasar
images. The values have been determined with a $\lambda=1$ bar, but
they are identical for $\lambda=0.5$ and 2 within $\pm0.01$. Each
table entry contains the value for $\nu=1.05$ and the differences to
the corresponding values for $\nu=1.15$ (upper index) and $0.95$
(lower index).}
\label{LocalLensingParameters}
\vspace{0.25cm}
\begin{tabular}{c|cc}
Image & $\kappa$ & $\gamma$ \\
\hline
A & $0.36^{-0.07}_{+0.07}$ & $0.40^{+0.03}_{-0.03}$ \\
B & $0.36^{-0.06}_{+0.07}$ & $0.42^{+0.03}_{-0.03}$ \\
C & $0.69^{-0.01}_{+0.01}$ & $0.71^{+0.09}_{-0.09}$ \\
D & $0.59^{-0.03}_{+0.02}$ & $0.61^{+0.06}_{-0.06}$ \\
\end{tabular}
\end{center}
\end{table} The values for $\nu=0.95$ and $1.05$ are similar to what one gets
for a circular $\nu=1$ power--law mass distribution (singular
isothermal sphere) with a mixture of internal and external shear \cite
{Kochanek1991,Witt1994}. It can be taken from tables~\ref
{ParametersofModels} and~\ref {LocalLensingParameters} that a change
of $\lambda$ does not cause a measurable change of observable
quantities except for the semi-minor axis of the bar; the uncertainty
of the model parameters is dominated by the uncertainty of the bulge
ellipticity $e$/exponent $\nu$.

\section{Properties of the bar}
\label{PropertiesOfTheBar}

The width of the bar has not been measured previously. In figure~1 of
their paper, Irwin {\etal} \shortcite {Irwin1989} present a contour
plot of the galaxy where the bar has been separated from the disk. In
this plot, the bar appears about 18 arcsec long, but only $\approx
2-4$ arcsec wide. It is just the less well-known minor axis that
enters the lensing model since the quasar images are situated in the
centre of the galaxy. In their image analysis Irwin {\etal} subtracted
structure from their image that is smooth on scales of $\approx 5-10$
arcsec. This procedure removed the galaxy disk very efficiently, but,
being long as well as thin, the bar shown in their figure~1 could also
be affected by this procedure.

A different approach to obtain the light distribution of the bar was
pursued by Schmidt~\shortcite {Schmidt1996}. Using a {\it Hubble Space
Telescope} ({\it HST}) I-band image taken with the Planetary Camera
prior to the first servicing mission \cite {Westphal1992}, the galaxy
was decomposed into bulge, disk and bar with analytical profiles that
have been convolved with the point-spread function of the
telescope. Exponential profiles \cite {Andredakis1994} were fitted to
bulge and disk. After these components were removed from the image,
the bar and the spiral arms remained. The bar light distribution could
be fit with a $\lambda=2$ Ferrers profile with $a=9.5\pm 1.0$ arcsec
and $b=1.0\pm 0.3$ arcsec. This light model can be combined with the
mass model for $\lambda=2$ from table~\ref {ParametersofModels} due to
the similar value for $b$; the I-band mass-to-light ratios of these
model components inside a circle of 0.9 arcsec are given by ${\cal
M}$/L$_{\rm I} \approx 4.8\, {\rm h}_{75}$ for bulge plus disk and
${\cal M}$/L$_{\rm I} \approx 5.0\, {\rm h}_{75}$ for the bar. The bar
mass detected with gravitational lensing and the bar light in the
I-band in this model both constitute a $5\%$ fraction of the total
mass respectively light inside $0.9$ arcsec.

This result is, however, dependent on the bulge model. For a de
Vaucouleurs--bulge \cite {deVaucouleurs1948}, a $\lambda=0.5$ Ferrers
profile fitted the bar light much better with a similar value for $a$,
but a much larger value $b=3.1\pm 0.9$ arcsec which cannot be combined
with the lensing model from table~\ref {ParametersofModels} since $b$
is very different. The unrefurbished {\it HST} point spread function
inhibited a clear distinction between the quality of fit of these
different bulge models, so that the question about the minor bar axis
is not decided yet.

A value of \mbox{${\rm b}\geq 1$ arcsec} could in connection with the
values for $b$ in table~\ref {ParametersofModels} be taken as evidence
in favour of steeper bar models, $\lambda \geq 2$. There is
observational evidence from the light distributions of real bars for
more boxy mass distributions \cite {Freeman1996}, so that it has to be
ascertained that this result is not a relic of the approximation of
the bar with an elliptical shape. At the positions of the quasar
images, however, an ellipse is a good approximation of a box due to
the large semi-major axis of the bar, $a\approx 9$~arcsec. A lensing
model with a steeper Ferrers profile with $\lambda=10$ leads to a
larger bar with $b=2.0$. This indicates that a measurement of the
extent of the bar could in principle constrain the mass distribution
of the bar. If one wants to use this method, high resolution images
with small seeing have to be used since an accuracy of the bar width
on the order of tenths of an arcsecond is needed.

\section{Discussion}
\label{Discussion}

In this paper we have presented a barred galaxy model for the
gravitational lens 2237+0305. We used a power--law elliptical mass
distribution for the bulge and chose the model for which the observed
ellipticity is predicted. It turned out that this model has an
exponent close to unity, which is compatible with other determinations
of lens mass profiles through gravitational lensing \cite
{Kochanek1995,Grogin1996}. The bar represents a small perturbation of
the deflection field of the bulge of the galaxy, amounting to $7.5\pm
1.5\times10^{8}\,{\rm h}_{75}^{-1}{\cal M}_{\sun}$ or about 5\%~of the
bulge mass in the critical region inside the quasar images.

The relative magnifications our model predicts for the quasar images
can be compared with the 3.6cm radio flux ratios published by Falco
{\etal} \shortcite {Falco1996}. Their results are also given in
table~\ref {ParametersofModels}. Falco {\etal} argue that it is
unlikely that the radio flux densities are variable from microlensing
due to the larger size of the radio emitting region as compared to the
optical continuum emitting region, although they cannot completely
rule out microlensing as an important effect in the radio. If
microlensing is not important, the model magnification ratios should
be identical to these measured flux ratios. It can be seen that only
the ratio $\mu_{\rm DA}$ between images D and A is not compatible with
their results although no effort was made to fit the flux ratios.

In order to find out more about the discrepancy of the ratio $\mu_{\rm
DA}$
between observation and model one has to make the relatively large
error bars from Falco {\etal} smaller through longer radio observation
of the object. Unfortunately, Q2237+0305 has a radio flux density of
only $\approx 1$~mJy \cite {Falco1996}, so that radio observations of
this object are very time-consuming; Falco {\etal} observed for 11
hours of which only five could be used eventually due to weather
conditions.

To get additional, independent arbiters for the model, it would be
very helpful to measure the time--delays in this system. Since the
time--delays are of the order of several hours, this has to
be done in a wavelength domain where the necessary intra-day
variability is likely to occur for a radio-quiet quasar, for example
in the x-ray regime as proposed by Wambsganss \& Paczy\'{n}ski
\shortcite {Wambsganss1994}. Also, monitoring in the radio would show
if the quasar image flux densities vary at these frequencies. Unless
we learn more about the radio flux densities and the time delays, it
is not possible to decide whether or not it is microlensing that
causes the low magnification of image D. In the optical, image D has
always been the faintest quasar image. In fact, in the first resolved
image of the quasar, image D was not visible at all \cite
{Tyson1985}. There is also spectroscopic evidence from optical data
that image D is undergoing demagnification \cite {Lewis1996}.

If image D is in fact microlensed in the radio, the consequences are
interesting. The scale size of the radio region could be less than the
characteristic scale of the caustic network. Alternatively, the radio
source could have an asymmetric structure like a jet that would have
differing microlensing properties for different paths of the
microlenses across the source.

Yet another way to significantly change the radio flux density of
image D would be a globular cluster or black hole \cite {Lacey1985}
with a mass of about $10^6 {\cal M_{\sun}}$ in the halo of the lensing
galaxy that is situated close to image D. An object of this mass would
magnify or demagnify the radio image of the quasar, depending on its
location with respect to the direction of the local shear. This
effect, the perturbation of lens models by $10^6 {\cal M_{\sun}}$
objects, has recently been treated by Mao \& Schneider \shortcite
{Mao1997}. With this, we can estimate that a surface mass density of
globular clusters or black holes of approximately $0.04\,\Sigma_{\rm
crit}$ ($\Sigma_{\rm crit}$ is defined in section~\ref {TheBulge}) or
$470\,{\rm h}_{75} {\cal M}_{\sun}/{\rm pc}^2$ is needed to
observe a demagnification of image D by $40\%$ or more with a
probability of 20\%. Higher surface mass densities would make it more
likely. This is much more than the globular cluster surface mass
density of about $1\,{\cal M_{\sun}}/$pc$^2$ seen in our Galaxy within
5~kpc of the Galactic centre \cite {Mao1997}. It thus seems unlikely
that the demagnification is due to a globular cluster.

The question of the existence of such a massive object near image D
could be solved with a method that was proposed by Wambsganss \&
Paczy\'{n}ski \shortcite {Wambsganss1992}. They showed that these
objects would bend or even create holes in the radio maps of
milliarcsecond jets of gravitationally lensed quasars. If we could
observe extended structure of the images of the radio-weak Q2237+0305,
features due to globular cluster or black hole lensing could be easily
identified because they would not be seen in the other gravitationally
lensed images of the quasar.

It has recently been suggested (Keeton, Kochanek \& Seljak 1996; Witt
\& Mao 1997) that a strong lensing perturbation is necessary for a
number of lenses in addition to an elliptical mass distribution in
order to model the observed image geometry. In our model for
Q2237+0305, the additional perturbation from the bar is small, but the
system is, nevertheless, unique in that we can see the perturbing
agent. For the lens systems mentioned in these papers, the better fit
for the models was obtained with an additional external shear. We saw,
when we compared our results with Wambsganss \& Paczy\'{n}ski
\shortcite {Wambsganss1994}, that the predictions for magnifications
or time--delays from shear models differ by up to a factor of two from
the predictions from elliptical mass deflectors. If one wants to get
lensing models with reliable predictions for magnifications or
time--delays, the necessary perturbations will ultimately have to be
generated in the models by mass components like dark matter haloes
\cite {Keeton1996} or bars.

\section*{Acknowledgments}

We thank Hans-J\"org Witt, Joachim Wambsgan{\ss} and the anonymous
referee for their comments on the manu\-script. RWS
gratefully acknowledges support by a Melbourne University Postgraduate
Scholarship, an Australian Overseas Postgraduate Research Scholarship
and by the Studienstiftung des deutschen Volkes. This research was
supported in part by the Deutsche Forschungsgemeinschaft (DFG) under
Gz. WA 1047/2-1.


\appendix

\section{Power--law elliptical mass distributions}
\label{PowerLawDeflection}

The lensing properties of the elliptical power--law mass distributions
from Equation~(\ref {SingularEllipticalProfiles}) with $\nu=1$ have
been described by Kassiola \& Kovner \shortcite {Kassiola1993} and
Kormann, Schneider, \& Bartelmann \shortcite {Kormann1994}. Kormann
{\etal} determined these by solving the Poisson equation for the
deflection potential $\psi$ (see Schneider {\etal} 1992 for the
definition of the deflection potential) in polar coordinates
$\theta=\sqrt {\theta_1^2+ \theta_2^2 }$ and $\varphi$

\begin{equation}
\frac{1}{\theta} \frac{\partial}{\partial \theta} \left( \theta
\frac{\partial \psi} {\partial \theta} \right) +\frac{1}{\theta^2}
\frac{\partial^2 \psi}{\partial \varphi^2} = 2\kappa \left(\theta,
\varphi \right).
\end{equation}

\noindent Their approach can be generalised for arbitrary real numbers
$\nu$ with $0\leq \nu <2$. Define $n=2-\nu$, $s_n= {\rm sin}\,
n\frac{\pi}{2}$, $c_n= {\rm cos}\, n\frac{\pi}{2}$ and 

\begin{equation}
\Delta \left( \varphi \right)= \sqrt {\frac{{\rm cos}^2 \varphi
}{\left( 1+ \epsilon\right)^2}+ \frac{{\rm sin}^2 \varphi }{\left( 1-
\epsilon \right)^2}}.
\end{equation}

\noindent Using the integrals
\begin{eqnarray}
{\cal N}_1 & = & \int_0^{\varphi} \frac{{\rm cos}\, n\varphi^{\prime}}
{\Delta^{\nu} \left( \varphi^{\prime} \right)} {\rm d}\varphi^{\prime}
\nonumber\\
{\cal N}_2 & = & \int_{\varphi}^{\frac{\pi}{2}} \frac{{\rm sin}\,
n\varphi^{\prime}} {\Delta^{\nu} \left( \varphi^{\prime} \right)} {\rm
d}\varphi^{\prime} \nonumber\\
{\cal N}_3 & = & {\cal N}_1 + \int_{\varphi}^{\frac{\pi}{2}}
\frac{{\rm cos}\, n\varphi^{\prime}} {\Delta^{\nu} \left(
\varphi^{\prime} \right)} {\rm d}\varphi^{\prime},
\end{eqnarray}

\noindent the deflection potential for $0\leq \varphi \leq
\frac{\pi}{2}$ is given by
\begin{eqnarray}
\label{psi_PowerLaw}
\lefteqn{\psi( \theta,\varphi )=} \nonumber\\
& & E_0\,\theta^n \! \left[ \frac{1}{n} {\cal N}_1\, {\rm
sin}\,n\varphi+ \frac{1}{n} \left( {\cal N}_2+ \frac{c_n}{s_n} {\cal
N}_3\ \right)\, {\rm cos}\, n\varphi \right]\!.
\end{eqnarray}

\noindent The potential for the other quadrants can be taken from this
result because of the elliptical symmetry of the mass
distribution. The deflection angles $\vec {\balpha}=\vec {\nabla}
\psi$ for this quadrant can be calculated from Equation~(\ref
{psi_PowerLaw}). The cartesian components are
\begin{eqnarray}
\lefteqn{\alpha_1( \theta,\varphi ) = E_0\, \theta^{n-1} \left[
\frac{}{} {\cal N}_1 \left( \frac{}{} {\rm sin}\,n\varphi\,{\rm cos}\,
\varphi- {\rm cos}\,n\varphi\, {\rm sin}\,\varphi \right) \right. }
\nonumber\\
& & \left. \mbox{} + \left( {\cal N}_2 + \frac{c_n}{s_n}\,{\cal N}_3
\right) \left( \frac{}{} {\rm cos}\,n\varphi\, {\rm cos}\,\varphi +
{\rm sin}\,n\varphi\, {\rm sin}\,\varphi \right) \right] \nonumber\\
\lefteqn{\alpha_2( \theta,\varphi ) = E_0\, \theta^{n-1} \left[
\frac{}{} {\cal N}_1 \left( \frac{}{} {\rm sin}\,n\varphi\,{\rm sin}\,
\varphi+ {\rm cos}\,n\varphi\, {\rm cos}\,\varphi \right) \right. }
\nonumber\\
& & \left. \mbox{} + \left( {\cal N}_2 + \frac{c_n}{s_n}\,{\cal N}_3
\right) \left( \frac{}{} {\rm cos}\,n\varphi\, {\rm sin}\,\varphi -
{\rm sin}\,n\varphi\, {\rm cos}\,\varphi \right) \right].
\end{eqnarray}
\noindent These expressions can be evaluated numerically. For
$\nu=n=1$ the integrals are analytically solvable and the formulas
become identical to the results by Kassiola \& Kovner \shortcite
{Kassiola1993} and Kormann {\etal} \shortcite {Kormann1994}. After
this work was completed, we discovered that Grogin \& Narayan
\shortcite {Grogin1996} also used power--law elliptical mass profiles
in their model of the lens of the double quasar 0957+561. In their
paper they present the deflection angles of power--law elliptical mass
distributions in a complex-valued lensing formalism in terms of the
complex hypergeometric function.

\section{Ferrers profiles}
\label {FerrersDeflection}

The deflection potential $\psi_0$ and the deflection angles $\vec
{\balpha}_0$ of an elliptical slice of constant surface density, a
Ferrers profile from Equation~(\ref {FerrersProfiles}) with $\lambda=0$,
are given by
\begin{eqnarray}
\psi_0 \left(\theta_1,\theta_2 \right) & = & -\frac{1}{2} ab
\kappa_{\rm c} \left(Q_{00}- \theta_1^2 Q_{10}- \theta_2^2 Q_{01}
\right)
\end{eqnarray}
\noindent and the cartesian components
\begin{eqnarray}
\alpha_{01} \left(\theta_1,\theta_2 \right) & = & ab\kappa_{\rm c}
\theta_1Q_{10} \nonumber\\
\alpha_{02} \left(\theta_1,\theta_2 \right) & = & ab\kappa_{\rm c}
\theta_2Q_{01},
\end{eqnarray}

\noindent where
\begin{eqnarray}
\label{q00}
Q_{00} & = & 2\,{\rm ln} \left( \sqrt{a^2+ \rho}+ \sqrt{b^2+ \rho}
\right)^{-1} \nonumber\\
Q_{01} & = & \frac{2}{a^2-b^2} \left( \sqrt{\frac{a^2+\rho}
{b^2+\rho}} - 1\right) \nonumber\\
Q_{10} & = & 2/\Delta \left( \rho \right)- Q_{01}.
\end{eqnarray}

\noindent $\rho$ is the positive solution of

\begin{equation}
\label{rho}
\frac{\theta_1^{2}}{a^{2}+\rho}+\frac{\theta_2^{2}}{b^{2}+\rho}
= 1
\end{equation}

\noindent outside the bar, and $\rho = 0$ inside the
bar. $\Delta\left( \rho \right)$ is given by

\begin{equation}
\Delta \left(\rho \right)= \sqrt {\left(a^2+\rho \right) \left(
b^2+\rho \right)}.
\end{equation}

\noindent $\vec {\balpha}_0$ has first been derived by Schramm
\shortcite {Schramm1990} by using the known force field for a
homogenous ellipsoid and letting the largest axis go to infinity in
order to get the corresponding 2-dimensional equations. To derive the
potential $\psi_0$ in terms of real-numbered coordinates $\theta_1,
\theta_2$, we applied Schramm's method to the results by Pfenniger
\shortcite {Pfenniger1984} for the potential and force fields of
3-dimensional Ferrers ellipsoids. This enables us to calculate
deflection potential and deflection angles for a whole family of
Ferrers profiles. The deflection potential for Ferrers
surface--density profiles with integer exponents $\lambda$ is given by
\begin{eqnarray}
\label{psi_mu}
\lefteqn{\psi_{\lambda} \left( \theta_1, \theta_2 \right)=} \nonumber\\
& & -\frac{ab\kappa_{\rm c}}{2(\lambda+1)}\, \int_{\rho}^{\infty}
\frac{ {\rm d}u} {\Delta(u)} {\left( 1-\frac {\theta_1^2}{a^2+u}-
\frac{\theta_2^2} {b^2+u} \right)^{\lambda+1 }}\!\!.
\end{eqnarray}

\noindent The corresponding 3-dimensional expression was first
de\-rived by Ferrers \shortcite {Ferrers1877}. For integers $j$, $k$,
$j\neq 0$ or $k\neq 0$, this integral can be split up into a sum
containing the coefficients

\begin{equation}
Q_{jk}= \int_{\rho}^{\infty} \frac{{\rm d} u}{\Delta \left(u
\right)} \frac{1}{\left(a^2+u \right)^j \left( b^2+u \right)^k}.
\end{equation}

\noindent For $Q_{00}$ use Equation~(\ref {q00}); a remaining
contribution from the infinite axis had to be subtracted
here. Pfenniger \shortcite {Pfenniger1984} solved his corresponding
integrals using recurrence relations. Translated into two dimensions,
the $Q_{jk}$ obey the relation

\begin{equation}
Q_{jk}= \left( Q_{j-1,k}- Q_{j,k-1} \right)/ \left(a^2-b^2 \right),
\end{equation}

\noindent as well as for $n>0$
\begin{eqnarray}
Q_{n0} & = &\frac{1}{2n-1}\left[ \frac{2}{\Delta \left( \rho \right)
\left(a^2+\rho \right)^{n-1}}- Q_{n-1,1} \right],\\
Q_{0n} & = &\frac{1}{2n-1}\left[ \frac{2}{\Delta \left( \rho \right)
\left(b^2+\rho \right)^{n-1}}- Q_{1,n-1} \right].
\end{eqnarray}

\noindent With these equations, the deflection potential
$\psi_{\lambda}$ as well as the deflection angles $\vec
{\balpha}_{\lambda}=\vec {\nabla} \psi_{\lambda}$ can be calculated
from Equation~(\ref {psi_mu}). The coefficients $Q_{ij}$ can be
treated as constant for the derivation with respect to $\theta_1$ and
$\theta_2$ since the definition of $\rho$ in Equation~(\ref {rho})
implies that $\frac {\partial \psi_{\lambda}} {\partial \rho}=0$.

\bsp

\label{LastPage}

\end{document}